\documentclass[twocolumn,amsmath,amssymb,superscriptaddress,nofootinbib]{revtex4-1}
\pdfoutput=1

\usepackage[english]{babel}
\usepackage{amsthm}
\usepackage{amssymb}
\usepackage{amsmath}
\usepackage[]{graphicx}
\usepackage{tensor}
\usepackage{color}
\usepackage{framed}
\usepackage[bookmarks,linktocpage, colorlinks=true, plainpages = false, citecolor = blue,  linkcolor=blue, urlcolor = blue, filecolor = blue]{hyperref} 
\usepackage{natbib}
\usepackage{dcolumn}
\usepackage{enumerate}
\usepackage{bm}
\usepackage{graphicx}
\usepackage{caption}
\usepackage{subcaption}

\theoremstyle{definition}

\graphicspath{{./figures/}}

\usepackage{tikz}
\usetikzlibrary{decorations.pathmorphing, patterns,shapes}

\input{epsf}

\begin{document}

\allowdisplaybreaks
\begin{titlepage}

\title{
A Minimal Explanation of the Primordial Cosmological Perturbations}
\author{Neil Turok}
\email{neil.turok@ed.ac.uk}
\affiliation{Higgs Centre for Theoretical Physics, James Clerk Maxwell Building, Edinburgh EH9 3FD, UK}
\affiliation{Perimeter Institute for Theoretical Physics, Waterloo, Ontario, Canada, N2L 2Y5}
\author{Latham Boyle}
\email{lboyle@perimeterinstitute.ca}
\affiliation{Perimeter Institute for Theoretical Physics, Waterloo, Ontario, Canada, N2L 2Y5}
\begin{abstract}
\vspace{.5cm}
\noindent 
We outline a new explanation for the primordial density perturbations in cosmology. Dimension zero fields are a minimal addition to the Standard Model of particle physics: if the Higgs doublet is emergent, they cancel the vacuum energy and both Weyl anomalies without introducing any new particles. Furthermore, the cancellation explains why there are three generations of elementary particles, including RH neutrinos.  We show how quantum zero point fluctuations of dimension zero fields seed nearly scale-invariant, Gaussian, adiabatic density perturbations.  We determine their amplitude in terms of Standard Model couplings and find it is consistent with observation. Subject to two simple theoretical assumptions, both the amplitude and the tilt we compute {\it ab initio} agree with the measured values inferred from large scale structure observations, with no free parameters.
\end{abstract}
\maketitle
\end{titlepage}


\section{Introduction}

Observations of the universe reveal a surprising simplicity. The vast array of new data ({\it e.g.}, Refs.~\cite{Planck:2018nkj, Planck:2018vyg, Gil-Marin:2014baa}) appears compatible with a vanilla $\Lambda$CDM model: as yet, no conclusive discrepancy has emerged.  Just three parameters -- the primordial baryon asymmetry, the dark matter density and the cosmological constant -- suffice to describe the energy content. Two more -- the primordial amplitude of scalar perturbations and a small, red spectral tilt -- account for large scale structure. We have recently proposed a new theoretical framework~\cite{Boyle:2018tzc, Boyle:2018rgh, Boyle:2021jej, Boyle:2022lyw, Boyle:2021jaz, Turok:2022fgq, Boyle:2022lcq}, capable of accounting for all features of the $\Lambda$CDM model, without requiring either inflation or any particles beyond those already present in the Standard Model (including RH neutrinos). Instead of postulating an early epoch of inflation, we suppose that the radiation-dominated epoch extended all the way back to the initial singularity. In this Letter, we present a new explanation for the primordial density perturbations and calculate their amplitude and spectral tilt in terms of Standard Model couplings.  The amplitude is directly compatible with the observations.  Under two simple theoretical assumptions, our {\it ab initio} calculations of both the amplitude and the tilt agree with the observations, with no free parameters.

Our basic hypothesis is that {\it the universe does not spontaneously break $CPT$ symmetry}~\cite{Boyle:2018tzc,Boyle:2018rgh}. We impose $CPT$ symmetry by analytically extending the spacetime to a ``mirror" universe on the other side of the bang, so that the extended spacetime possesses a natural time reversal isometry. Within classical GR, such an extension is possible if the material content of the universe is Weyl (or locally conformal) invariant so that, in effect the matter does not ``see" space shrinking away as we go back to the bang. In fact, it is well known that there is a class of regular solutions to the Einstein-relativistic perfect fluid equations, where the fluid has a traceless stress tensor, {\it i.e.}, $P={1\over 3} \rho$, and the line element takes the local form $ds_E^2=\eta^2\left(-d\eta^2+h_{ij}(\eta,x )dx^i dx^j\right)$, with $h_{ij}(\eta,x )=h^0_{ij}(x)+h^2_{ij}(x)\eta^2 +\dots$ near $\eta=0$ and $h^0_{ij}(x)$ an arbitrary, regular spatial metric~\cite{Belinsky:1982pk,Newman:1992tc,Newman:1992cx}. Such a singularity is purely {\it conformal}, {\it i.e.}, $g_{\mu \nu}(x) = \Omega^2(x) \hat{g}_{\mu \nu}(x) $, with $ \hat{g}_{\mu \nu}$ the metric on $\hat{\cal M}$, a regular spacetime. The conformal factor $\Omega (x)$ smoothly changes sign across a 3-surface $\hat{\Sigma} \in \hat{{\cal M}}$, dividing $\hat{\cal M}$ into two halves.  The above solutions have an isometry $T:\eta\rightarrow -\eta$ which protects the conformal character of the singularity: perturbations with this symmetry obey reflecting boundary conditions at $\eta=0$ and remain regular there~\cite{Boyle:2018tzc}.  If the extended cosmology is described by a gravitational path integral whose initial and final states are related via $CPT$, such metrics appear to be legitimate saddles~\cite{footnote}.  More ``generic" GR solutions with Kasner-type singularities or BKL chaos~\cite{BKL} would  be excluded because they are singular and therefore not genuine stationary points of the action. Interference between such paths would cause them to cancel out of the path integral. Thus, it does not seem unreasonable to suppose that the only $CPT$-symmetric cosmological saddles are spacetimes with a conformal singularity, but nothing worse~\cite{Boyle:2021jej,Boyle:2022lyw}.

Penrose has long emphasized~\cite{Penrose} that the big bang seems to have been a special type of singularity, quite different from the more generic type found inside black holes. Metrics of the above form are analytic functions of the conformal time, meaning that all curvature invariants are meromorphic and all Weyl-invariant quantities are finite. In Refs.~\cite{Boyle:2018tzc, Boyle:2018rgh, Boyle:2021jej, Boyle:2022lyw, Boyle:2021jaz, Turok:2022fgq, Boyle:2022lcq}, our approach has been to take seriously the cosmological solutions of the Einstein equation, including their maximal analytic extensions, both along the real time axis and more generally in the complex time plane.  Even if the Einstein equations are no longer valid in the immediate vicinity of the bang, {\it i.e.}, in the Planckian regime, the meromorphicity of the spacetime and the fields living on it allows us to reason about them without ever coming too close to the singular point. 

Our recent calculation of the gravitational entropy for cosmological spacetimes lends strong support this approach. We find that homogeneous, isotropic and flat universes like ours, with a small, positive cosmological constant, are entropically favoured~\cite{Turok:2022fgq,Boyle:2022lcq}. Invoking $CPT$-symmetry also makes viable a remarkably simple candidate for the cosmic dark matter: a stable, right-handed (RH) neutrino \cite{Boyle:2018tzc, Boyle:2018rgh}. Although such a particle was never in thermal equilibrium, imposing $CPT$ symmetry on its vacuum state fixes its abundance today. Matching the dark matter density then fixes its mass to be $\approx 4.8\times 10^8$ GeV. The stability of this RH neutrino implies that one of the LH neutrinos is massless, a prediction which will be closely tested by galaxy surveys in the next 3-5 years~\cite{Chen:2021vba,eBOSS:2020yzd}.

\section{Dimension zero fields}

If the big bang singularity is only conformal then, for Weyl-invariant matter, the matter evolution is well-defined throughout the extended spacetime. Therefore, a crucial question is whether the Standard Model is in fact consistent with Weyl symmetry at high temperatures, densities and spacetime curvatures. 
There are at least two obstacles -- two contributions to the trace anomaly, associated with the violation of local and global scale invariance, respectively.  First, even at the level of free fields on curved spacetime, we have the ``local" contribution, the Weyl anomaly \cite{Duff:1993wm}.  Second, when interactions are included, we have the ``global" contribution, arising from the running couplings in the Standard Model.  We proceed by cancelling both types of anomaly at leading order and high temperatures. 

A dimension zero scalar has classical action $S_0[\varphi]=-\frac{1}{2}\int d^{4}x\sqrt{-g}\,\varphi\Delta_{4}\varphi$ with $\Delta_{4}$ the Paneitz operator~\cite{Fradkin:1981jc, Fradkin:1982xc, Paneitz}. Although this action is classically Weyl-invariant (because $\sqrt{-g} \Delta_{4}$ is), the stress tensor for $\varphi$ has a quantum anomalous trace called the Weyl anomaly,
\begin{equation}
\langle T^{\varphi \lambda}_{\lambda} \rangle=a (E-\frac{2}{3}\Box R)+c \,C^{2}+\xi \Box R,
\label{ea1}
 \end{equation}
where $E$ is the Euler density, $R$ the Ricci scalar and $C$ the Weyl curvature tensor, $a={7\over 90 (4 \pi)^2}$, $c=-{1\over 15 (4 \pi)^2}$ and $\xi={16\over 135 (4 \pi)^2}$~\cite{Fradkin:1983tg}.  We recently pointed out that adding $N_0=36$ such fields to the Standard Model precisely cancels the ``local" anomalies proportional to $E-\frac{2}{3}\Box R$ and $C^2$, provided the Higgs doublet is not a fundamental field. (The $\Box R$ contribution may be cancelled by adding a local $R^2$ counterterm to the action, so we shall omit it in the discussion below.) Remarkably, the same 36 dimension zero fields also cancel the vacuum energy in the SM, at free field order, iff there are 3 generations of fermions~\cite{Boyle:2021jaz}. 
 
 The trace anomaly contributed by a dimension zero field is reproduced by an effective action $\Gamma^\varphi = -\frac{1}{2}\int  a \,\varphi \Delta_{4}\varphi +\Delta \Gamma^\varphi$, with
 \begin{equation}
 \Delta \Gamma^\varphi=\frac{1}{2} \int  \Big[a (E-\frac{2}{3}\Box R)+c  C^{2}+ d  T_\beta^{SM} \Big]\varphi,
 \label{ea2}
 \end{equation}
For later convenience, we rescaled $\varphi$ to put a factor of $a$ in front of the kinetic term. The $a$ and $c$ terms in (\ref{ea2}) reproduce the trace anomaly (\ref{ea1}), because the $(E-\frac{2}{3}\Box R)$ term in (\ref{ea2}) is {\it not} Weyl invariant and thus contributes to $T^{\varphi \lambda}_{\lambda}$, as we shall see. In the same vein, we couple $\varphi$ to the Standard Model's high temperature quantum trace anomaly $T_\beta^{SM}$ in order to cancel the ``global" part of the trace anomaly, arising from running couplings. To see how all this works, calculate the stress tensor from (\ref{ea2}):
 \begin{eqnarray}
 &&T^{\varphi \lambda}_{\,\lambda}= - {2  g^{\alpha \beta} \over \sqrt{-g}}{\delta \Delta \Gamma^\varphi \over \delta g^{\alpha \beta}}=2 a \Delta_{4}\varphi \cr
 &&= a(E-\frac{2}{3}\Box R)+c \,C^{2}+ d \,T_\beta^{SM}.
 \label{ea3}
 \end{eqnarray}
where we used the  equation of motion for the $\varphi$. The global anomaly is cancelled if $d=-1/N_0=-{1\over 36}$ (we assume the dimension zero fields couple identically to $T_\beta^{SM}$). Using a classically non-Weyl invariant term to cancel a quantum trace anomaly in this way is a familiar procedure in the sigma model approach to string theory, a 2d model of quantum gravity (for a review, see \cite{Callan:1989nz}).

\section{Trace anomaly in the SM at High T}

The Standard Model fails to be Weyl-invariant at high temperature, due to its running couplings. From standard thermodynamic relations, the thermal average of the stress tensor's trace, $T_\beta \equiv \langle T^{\,\,\lambda}_{\lambda} \rangle_\beta= 3P-\rho=T (dF/dT)-4 F$, with $F(T)$ the free energy density. For the Standard Model, $F(T)$ has been calculated using a combination of perturbative and non-perturbative methods~\cite{Parwani:1994je,Arnold:1994eb,Andersen:1995ej,Andersen:2009tw,Gynther:2005dj,Kajantie:2002wa,Hietanen:2008tv}. A term $\propto T^4$ makes no contribution to $T_\beta$, but a term $A T^4 \ln (T/\Lambda)$, with $\Lambda$ some fixed renormalization scale, contributes $ A T^4$. The leading perturbative contributions to $T^{SM}_\beta$ are (see Ref.~\cite{Arnold:1994eb}, Eq.~(5.1))
\begin{equation}
T_\beta^{SM} =\left({125\over 108}\alpha_Y^2 -{95\over 72} \alpha_2^2 -{49\over 6} \alpha_3^2\right) T^4 \equiv c^{SM}_\beta T^4 
 \label{ea4}
\end{equation} 
where $\alpha_{Y,2,3} \equiv g_{Y,2,3}^2/(4 \pi) $. In their notation, for $U(1)_Y$, $d_A$=1, $C_A=0$, $S_F=\sum_i Y_i^2=5$. For $SU(2)_L$, $d_A$=3, $C_A=2$, $S_F=3$ and for $SU(3)$, $d_A=8$, $C_A=3$, $S_F=3$.  

The action for a relativistic fluid is $S_{F}=-\int d^4 x \sqrt{-g} \rho(x)$ where $\rho(x)$ is its energy density~\cite{Fockbook}. In local thermal equilibrium, and to lowest order in couplings, the density is related to the temperature by $\rho = {\pi^2\over 30} {\cal N}_{eff} T^4$ where ${\cal N}_{eff}$ is the effective number of relativistic degrees of freedom. For the Standard Model's gauge fields and 3 generations of fermions, including RH neutrinos, ${\cal N}_{eff}=108$. If one RH neutrino is stable dark matter, and thus decoupled from the hot plasma, ${\cal N}_{eff}$ is reduced by $7/4$. Similarly, including the Higgs doublet would raise ${\cal N}_{eff}$ by 4.  Combining $S_{F}$ with the last term in (\ref{ea2}) and re-expressing the temperature in terms of the density, the effective action for the fluid, including its coupling to the dimension zero fields, becomes
\begin{equation}
S_{F,\varphi}=-\int d^4x \sqrt{-g} \rho(x) (1+c_\chi \chi); \quad c_\chi= {15 \, c^{SM}_\beta \over \pi^2 {\cal N}_{eff} },
\label{ea5}
\end{equation}
where $\chi \equiv N_0^{-1} \sum_{j=1}^{N_0} \varphi_j$.

\section{Long wavelength perturbations}
 
We now combine the fluid action (\ref{ea5}) with the Einstein action for gravity in order to study the influence of the scalar fluctuations $\delta \varphi$ on long wavelength modes in the early universe.  The comoving modes of interest for large scale structure today were far outside the Hubble horizon at early times, {\it i.e.}, deep in the radiation era. Provided their amplitude is sufficiently small, their evolution can be accurately tracked in a local approximation where derivatives are subdominant. The quadratic action for the fields $\varphi$, which determines their quantum fluctuations, is Weyl invariant so it makes sense to work in conformal coordinates for the background: $ds^2\approx a(\eta)^2(-d\eta^2+d {\bf x}^2)$. The linear coupling of the $\varphi$ to $T_\beta$ drives a classical background solution $\varphi_C$ but, since the action is only quadratic, writing $\varphi=\varphi_C+\delta \varphi$ the fluctuations $\delta \varphi$ completely decouple from $\varphi_C$.  In the vacuum, which is their only physical state~\cite{Bogoliubov,Rivelles:2003jd}, they possess scale-invariant equal-time correlations~\cite{Boyle:2021jaz}
 \begin{equation}
  \langle\delta \varphi_i(\eta,{\bf x})\delta \varphi_j(\eta,{\bf x}')\rangle= \delta_{ij}  \int\frac{d^{3}{\bf k}}{(2\pi)^{3}}\frac{{\rm e}^{i{\bf k}\cdot({\bf x}-{\bf x}')}}{4 a k^{3}},
\label{ea6}
\end{equation}
where we use the same normalization as in (\ref{ea2}). The linear combination $\delta \chi \equiv N_0^{-1} \sum_{j=1}^{N_0} \delta \varphi_j$  has the same correlator but with $\delta_{ij}$ replaced by $1/N_0$. 
Since only $\chi$ appears in the action (\ref{ea5}), it is the only combination appearing at linear order in cosmological perturbation theory, and the resulting perturbations will be adiabatic. 

The comoving modes of observational interest today have wavelengths far outside the Hubble horizon at early times, so  it is reasonable in a first approximation to ignore spatial derivatives. Likewise, for wavelengths outside the Hubble horizon, it is reasonable to ignore the time evolution of $\delta \chi$. Writing the perturbed line element as  $ds^2\approx a(\eta,{\bf x})^2(-d\eta^2+d {\bf x}^2)$, the  $\sqrt{-g} G^{\,\,\eta}_\eta$ Einstein equation, {\it i.e.}, the Friedmann equation, becomes
 \begin{equation}
(\partial_\eta a)^2  \approx {8 \pi G\over 3} \rho \,a^4 (1+c_\chi \delta \chi({\bf x})).
\label{ea7}
\end{equation}
Since the background quantity $\rho \,a^4$ is nearly constant in the radiation era near the bang, and assuming $c_\chi \delta \chi({\bf x})\ll 1$, the approximate solution is  $ a(\eta,{\bf x}) \approx c_r \eta (1+ {1\over 2}c_\chi \delta \chi({\bf x})) $, with $c_r$ a constant. Thus the solution is a conformal perturbation of the four dimensional cosmological line element. In terms of proper time for the background, $t\equiv \int_0^\eta c_r \eta d\eta = {1\over 2} c_r \eta^2$,  we obtain 
 \begin{equation}
ds^2\approx (1+2 \sigma({\bf x}))\big(-dt^2+a(t)^2 d{\bf x}^2\big), \label{ea9}
\end{equation}
with $\sigma({\bf x}) = {1\over 2} c_\chi \delta \chi({\bf x})$. 

The comoving curvature perturbation ${\cal R}$ may be found by gauge transforming to a comoving, synchronous gauge 
\begin{equation}
ds^2\approx -d t_s^2 + a(t_s)^2 \big(1+2  \alpha({\bf x})\big) d{\bf x}^2.\label{ea10}
\end{equation}
For long wavelengths, $\alpha({\bf x})$ is a local dilation of space which is conserved by the Einstein equations (see, {\it e.g.}, \cite{Lyth:2004gb}).  The comoving curvature ${\cal R}$ is defined to be the coordinate invariant quantity which equals $\alpha$ in such a gauge. (The Ricci curvature of the line element $\big(1+2 \alpha({\bf x})\big) d{\bf x}^2$ is $ R^{(3)}\approx -4 {\bf \nabla}^2\alpha$, to linear order in $\alpha$.)

We transform the line element (\ref{ea9}) into the form (\ref{ea10}) as follows. First, we define $t_s=t+\int_0^t dt \sigma$. Next, we identify $a(t)(1+\sigma)$ with $ a(t_s)(1+\alpha)\approx a(t)(1+ H\int \sigma +\alpha)$, where $H=(da/dt)/a$. Thus, we obtain $\alpha=\sigma-H\int \sigma$. Dividing by $H$, differentiating, and using the background equation $dH/dt= -{3\over 2} H^2 (1+w)$ with $w\equiv P/\rho$, we find 
\begin{equation}
{\cal R}\equiv\alpha \approx \sigma- {2 (\sigma-H^{-1} \partial_t{\sigma})\over 3(1+w)}. \label{ea11}
\end{equation}
For $w={1\over 3}$ and $\partial_t {\sigma}$ small, as we expect for long wavelengths, ${\cal R}\approx {1\over 2} \sigma =  {1\over 4} c_\chi \delta \chi({\bf x})$. This is our prediction for the long-wavelength, comoving curvature perturbation which will later seed large scale structure formation. 

It is straightforward to show that $\alpha$ is equivalent to the usual expression for $ {\cal R}$ derived in Newtonian gauge, where the line element is $-dt_N^2(1+2 \Phi) +a(t_N)^2 (1-2 \Phi)d{\bf x}^2$. Starting from (\ref{ea10}), we pass to Newtonian gauge by setting $dt_s\approx dt_N (1+\Phi)$ and $a(t_s)(1+\alpha)\approx a(t_N)(1-\Phi)$, to first order in $\Phi$. Synchronizing $t_N$ and $t_s$ at zero, we find $t_N\approx t_s-\int_0^{t} d t \Phi$ and $\alpha\approx -\Phi-H\int_0^t dt \Phi$. Dividing this last equation by $H$ and using $\partial_t\alpha\approx 0$ yields
\begin{equation}
{\cal R} \approx -\Phi -{2(\Phi+H^{-1} \partial_t \Phi)\over 3(1+w)}, \label{ea12}
\end{equation}
the usual definition~\cite{Mukhanov:1990me}. For superhorizon adiabatic modes in the matter era ($w=0$),  $\Phi$ is nearly constant so ${\cal R}\approx -{5\over 3} \Phi$, a result we shall use later. 

Note that this description of the evolution of long wavelength modes is insensitive to higher derivative terms such as may arise for the fluid ({\it e.g.}, viscosity), for gravity (higher powers of the curvature), or from the effective action for the dimension zero fields. The only requirements are that i) the perturbations are adiabatic and ii) the total stress tensor is conserved (see, {\it e.g.}, \cite{Lyth:2004gb}).

\section{The perturbation amplitude}

We now combine the results of previous sections. From ${\cal R}\approx{1\over 4} c_\chi \delta \chi({\bf x})$, given below (\ref{ea11}), with $ c_\chi$ from (\ref{ea5}) and the $\delta \chi$ correlator explained below (\ref{ea6}), we find
\begin{equation}
\langle |{\cal R}_{\bf k}|^2 \rangle = {c_\chi^2  \over 64 k^3 a N_0}.  \label{ea14}
\end{equation}
The power spectrum of primordial perturbations is usually presented in terms of ${\cal P}_{\cal R}\equiv \langle |{\cal R}_{\bf k}|^2 \rangle k^3/(2 \pi^2)$, so that the real-space 
variance $\langle {\cal R}^2 \rangle=\int {\cal P}_{\cal R} d \ln k$, {\it i.e}, ${\cal P}_{\cal R}$ is the variance per logarithmic interval in comoving wavenumber $k$.  From (\ref{ea14}), with $N_0=36$, we obtain
\begin{equation}
 {\cal P}_{\cal R}= {3^2 5^3  \over 7(2\pi)^4} \left( {c^{SM}_\beta \over {\cal N}_{eff} } \right)^2,  \label{ea15}
\end{equation}
where $c^{SM}_\beta$ is given in (\ref{ea4}) and, with one stable RH neutrino and no Higgs at the Planck scale, ${\cal N}_{eff}=106{1\over 4}$.

\section{The spectral tilt }

The second important parameter characterizing the large scale primordial perturbations is the spectral tilt. Namely, observations indicate that ${\cal P}_{\cal R}$ has a modest dependence on comoving wavenumber $k$: modelling this as a power law, one finds
\begin{equation}
 {\cal P}_{\cal R}(k)\approx {\cal P}_{\cal R}(k_*) (k/k_*)^{n_s-1},
 \label{ea16}
\end{equation}
with $n_s-1\ll 1$ and $k_*$ a convenient pivot scale. 

It is not  hard to see how a tilt might arise, in our derivation. The constant $c_\beta$ is dominated by the fourth power of the $SU(3)$ coupling, which increases with distance. Hence one naturally expects more power on long wavelengths, {\it i.e.}, a red spectral tilt. We shall now give a heuristic analysis which, as it turns out, gives a remarkably accurate match to the observations. However,  we caution the reader that our result rests on some key theoretical assumptions which we have yet to verify. 

To understand the scale dependence of $c_\beta$, we return to a more microscopic picture of the global Weyl anomaly. We focus on the dominant, $SU(3)$ contribution and leave the number of flavors $n_f$ as a free parameter.  We proceed by calculating the high temperature trace $T_\beta$  in two steps. As is well known, a heuristic derivation of the trace anomaly is, first, to rescale the vector potential to re-express the action $\int (-{1\over 4}F^2)$ as $\int \left(-{1\over 4} \tilde{F}^2/g^2\right)$. The only dependence of the action on scale is then through the factor $1/g^2$. Taking $\mu \partial_\mu$ of the action where $\mu$ is a mass scale, one obtains $\int\left( {1\over 4} F^2  \beta_\alpha /\alpha \right)$, where $\beta_\alpha\equiv \mu \partial_{\mu} \alpha$. By relating this to the usual derivation of the stress tensor, one finds $T^{\,\,\lambda}_{\lambda}= {1\over 4} F^2 \beta_\alpha /\alpha$. For QCD with $n_f$ flavors ($n_f=6$ in the SM), $\beta_\alpha =-(11-{2\over 3} n_f) (\alpha^2/2 \pi)$. Let us now interpret the high temperature anomaly, generalized to $n_f$ flavors, in the light of this result. From \cite{Arnold:1994eb},  $T_\beta=-(12+5 n_f)(11-{2\over 3} n_f)  \alpha^2 T^4/36$, where $\alpha=\alpha_3$, the QCD coupling. The result clearly factorizes, with the second factor proportional to $\beta/\alpha$. The first is accounted for if $\langle F^2\rangle_{T} \sim {2 \pi\over 9} \alpha (12+5 n_f) T^4$. This interpretation is consistent with the fact that, for free gauge fields, $\langle F^2\rangle_T \propto \langle E^2-B^2\rangle_T=0$ by equipartition. $\langle F^2\rangle_T$ is only nonzero due to interactions: because the theory is invariant under $g\rightarrow -g$, the lowest allowed order is $g^2$. 

From this discussion, it is clear that the two factors of $\alpha$ appearing in $T_\beta$ have different origins. The first factor (coming from $\langle F^2\rangle_T$) is due to interactions in the plasma, and is a background quantity which should be calculated at the high temperature scale $T$. However, the second is associated with the running of $\alpha$. It is plausible that this factor depends on the spatial wavelength at which $\delta \chi $ couples to the plasma. This being the case, ${\cal P}_{\cal R}(k)$ runs with scale as {\it two} powers of $\alpha$, as we change the spatial wavelength at a fixed time (or temperature) so that (using $\beta_\alpha$ for QCD with $n_f=6$)
\begin{equation}
  n_{s}-1=\frac{d\,{\rm ln}\,{\cal P}_{{\cal R}}(k)}{d\,{\rm ln}\,k}=\frac{2}{\alpha}\beta_{\alpha}\approx-\frac{7\alpha_{3}}{\pi},
  \label{ea17}
\end{equation}
to leading order.  We now make our final, key assumption. It seems most natural that the expressions we have derived above, involving $\alpha_{3}$ evaluated at the Planck scale, correctly describe the amplitude and spectral tilt on comoving spatial scales comparable to the Planck length at the Planck time.  Thus, $n_s$ represents the scaling dimension of the operator ${\cal R}$. If, at high temperature, $n_s$ is a constant with respect to length scale (as it would be, {\it e.g.},  if ${\cal R}$ is an operator in a conformal field theory) then we can use it to extrapolate our results to arbitrarily large comoving scales, including those observed today.

\section{Comparison with Observation }

Let us compare these predictions with observation. Following the above discussion, we regard the prediction for $ {\cal P}_{\cal R}$ in (\ref{ea15}), with $c_\beta$ evaluated at the Planck scale, as being valid on a comoving length scale of the Planck length at the Planck temperature, about one mm today. Using the gauge coupling constants at the Planck scale,  $\alpha_Y=0.0181;\quad \alpha_2=0.0203;\quad \alpha_3=0.0189$, from \cite{Buttazzo:2013uya}, and (\ref{ea4}), we find $c_\beta=-0.0031$. The $SU(3)$ contribution comprises  $95\%$ of the total. From (\ref{ea17}),  we infer
\begin{equation}
  n_{s}\approx 1-7\alpha_{3}(M_P)/\pi=0.958,
  \label{ea18}
\end{equation}
compared to the Planck satellite's estimate $n_s=0.9587\pm 0.0056$~\cite{Planck:2018jri} when the first and second logarithmic derivatives of $n_s$ are treated as additional parameters (neither are found to be significantly different from zero). 

We then use (\ref{ea15}) and (\ref{ea16}) to express our prediction for ${\cal P}_{\cal R}(k)$ in terms of the pivot scale used by the Planck satellite, $k_*=0.05$ Mpc$^{-1}$ today~\cite{Planck:2018nkj}, and the comoving wavenumber $k_P\approx 2 \pi/(1\,{\rm mm})$ today, corresponding to the Planck scale at the Planck time:
\begin{equation}
 {\cal P}_{\cal R}(k)= {\cal P}_{\cal R} \left({k_*\over k_P}\right)^{n_s-1} \left({k\over k_*}\right)^{n_s-1} \equiv A \left({k\over k_*}\right)^{n_s-1}.
 \label{ea19}
\end{equation}
From the Planck satellite's estimate of $n_s$, we obtain  $(k_*/ k_P)^{n_s-1}=14.8\pm 5.1$. From (\ref{ea19}), we then predict $A=12.9\pm 4.5\times 10^{-10}$. Planck's measured value $A=21\pm 0.3 \times 10^{-10}$~\cite{Planck:2018vyg} is less than 2$\sigma$ away. (To intuitively estimate the associated CMB anisotropy, see \cite{footnote2}). 

The agreement between the observations and our calculated values for both the scalar amplitude and the spectral tilt is remarkable, all the more so in view of the significant remaining uncertainties. Our result is based on an extrapolation all the way back to the Planck time, and over more than 30 orders of magnitude in comoving length scale. As we indicated, if $n_s$ represents an approximately constant scaling dimension that vast extrapolation would nevertheless be justified. Here, there is potential overlap between our ideas and those of asymptotic safety, for example~\cite{Reuter:1996cp,Bonanno:2020bil,Eichhorn:2022gku}.  Second, we only included the leading order coupling constant dependence: this should clearly be extended to higher orders. Third, we neglected the influence on dimension zero field correlations of the global structure of cosmological instantons~\cite{Boyle:2022lcq} as well as potential boundary effects~\cite{Chalabi:2022qit} at the bang. Finally, we have not discussed the evolution of the dimension zero fields beyond linear order or at late times. Presumably, they play a role in the emergence and condensation of the Higgs field, for example. While the results of this paper are very encouraging, much remains to be understood. 

{\bf Acknowledgements:}  We thank Sam Bateman, Astrid Eichhorn, Steven Gratton, Chris Herzog, Slava Mukhanov, Malcolm Perry, Andreas Stergiou, Kostas Tsanavaris and Roman Zwicky for helpful discussions. The work of NT is supported by the STFC Consolidated Grant `Particle Physics at the Higgs Centre' and by the Higgs Chair. Research at Perimeter Institute is supported by the Government of Canada, through Innovation, Science and Economic Development, Canada and by the Province of Ontario through the Ministry of Research, Innovation and Science.


\begin{thebibliography}{99}
\bibitem{footnote} Treating the quantum fluctuations about such a classical background correctly is still a delicate matter because they may become strongly coupled near the bang.
\bibitem{footnote2}
In rough magnitude, the large scale CMB anisotropy is given by the Sachs-Wolfe formula $\delta T/T ={1\over 3} \Phi \approx-{1\over 5} {\cal R}$ in the matter era (see below Eq.~(\ref{ea12})). From the Planck satellite's measured value for $A$ it follows that $(\delta T/T)_{rms} \approx {1\over 5} {\cal P}_{\cal R}^{1\over 2}\approx 10^{-5}$ on large scales.
\end{thebibliography}
\end{document}